%% file: main.tex
\documentclass[sigconf]{acmart}

\usepackage{booktabs}
\usepackage{graphicx}
\usepackage{subcaption}
\usepackage{adjustbox}

\setcopyright{none}

\renewcommand\footnotetextcopyrightpermission[1]{}
\begin{document}
\title{Sequeval: A Framework to Assess and Benchmark Sequence-based Recommender Systems}

\author{Diego Monti}
\orcid{0000-0002-3821-5379}
\affiliation{Politecnico di Torino}
\email{diego.monti@polito.it}

\author{Enrico Palumbo}
\orcid{0000-0003-3898-7480}
\affiliation{ISMB-EURECOM}
\email{palumbo@ismb.it}

\author{Giuseppe Rizzo}
\orcid{0000-0003-0083-813X}
\affiliation{ISMB}
\email{giuseppe.rizzo@ismb.it}

\author{Maurizio Morisio}
\affiliation{Politecnico di Torino}
\email{maurizio.morisio@polito.it}

\begin{abstract}
In this paper, we present \texttt{sequeval}, a software tool capable of performing the offline evaluation of a recommender system designed to suggest a sequence of items. A sequence-based recommender is trained considering the sequences already available in the system and its purpose is to generate a personalized sequence starting from an initial seed. This tool automatically evaluates the sequence-based recommender considering a comprehensive set of eight different metrics adapted to the sequential scenario. \texttt{sequeval} has been developed following the best practices of software extensibility. For this reason, it is possible to easily integrate and evaluate novel recommendation techniques. \texttt{sequeval} is publicly available as an open source tool and it aims to become a focal point for the community to assess sequence-based recommender systems.
\end{abstract}

\maketitle

\input{sections/01_introduction.tex}
\input{sections/02_architecture.tex}
\input{sections/03_usage.tex}
\input{sections/04_conclusion.tex}

\bibliographystyle{ACM-Reference-Format}
\bibliography{references} 

\end{document}

%% file: sections/01_introduction.tex
\section{Introduction}

Traditional Recommender Systems (RSs) usually do not consider the temporal dimension of user preferences when suggesting a set of items. This simplifying hypothesis may represent a limitation in domains characterized by the rapid consumption of different items one after the other. Even if some authors proposed RSs capable of considering training items as sequences~\cite{He2017}, the idea of suggesting sequences of items instead of lists ranked by relevance is not widespread. On the other hand, this problem is quite similar to the task of generating a phrase given its initial words~\cite{Jurafsky2008}.

Consider, for example, the context of musical streaming services. Starting from the human-curated playlists already available in the system, a sequence-based RS should be capable of creating a personalized playlist for the target user given an initial seed. The initial seed may be a song, a set of songs, or a genre.

To the best of our knowledge, there is no standardized protocol for performing an offline experiment with sequence-based RSs. The lack of standardized protocols and metrics is a well-known problem inside the RS evaluation community, as it leads to research results that are difficult to compare~\cite{Jannach2015}.

Several software tools designed to simplify the development of novel recommendation algorithms are available. Such tools usually also provide a module devoted to their evaluation, but the results obtained are often not comparable because of the differences in how the protocols and the metrics are implemented~\cite{Said2014}.

In this paper, we present \texttt{sequeval}, a Python implementation of an evaluation framework designed for comparing sequence-based RSs. This software package is freely available on GitHub\footnote{\url{https://github.com/D2KLab/sequeval}} and its usage is described in an introductory video.\footnote{\url{https://doi.org/10.6084/m9.figshare.6854726}}

%% file: sections/02_architecture.tex
\section{System Architecture}

\begin{figure}
\includegraphics[width=\linewidth]{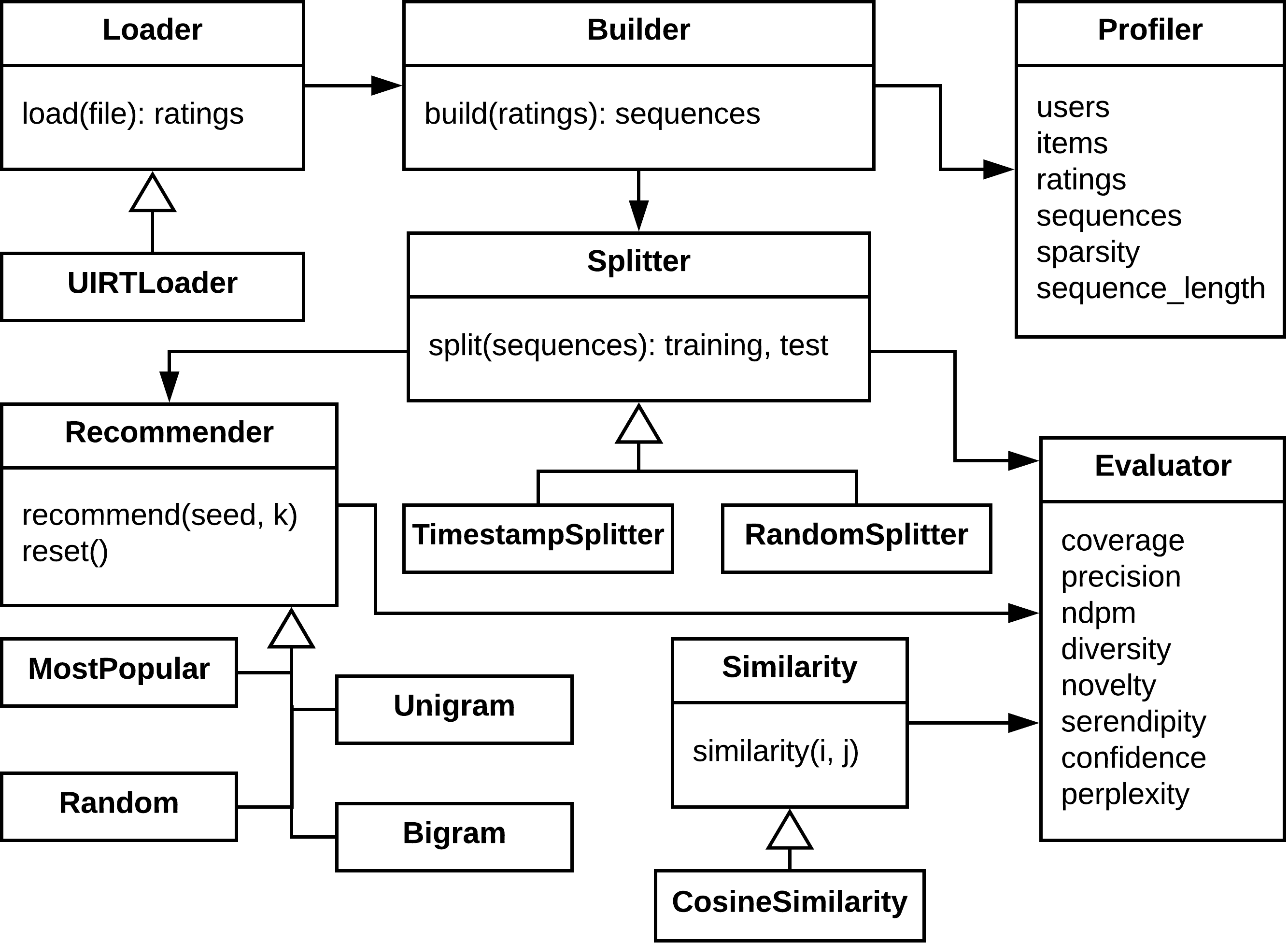}
\caption{Simplified UML class diagram}
\label{fig:uml}
\end{figure}

\texttt{sequeval} is a Python package composed of different modules which are graphically represented in Figure~\ref{fig:uml}. For each component, we defined an abstract class and then we realize one or more possible implementations in order to enable extensibility.

The \textit{loader} module is in charge of reading an input file containing user ratings. We have implemented a concrete \textit{loader} capable of processing a textual file in a MovieLens-like format (UIRT), but the support to other formats can be easily added. It is optionally possible to ignore users or items that do not have a minimum number of ratings, in order to avoid data sparsity issues.

The \textit{builder} module creates the sequences of items from the initial ratings: ratings from the same user that are distant in time less then a threshold are grouped inside the same sequence. Ratings that do not belong to any sequence are discarded.

The \textit{profiler} module computes some statistics about the generated sequences, for example their average length.

The \textit{splitter} module assigns the sequences created by the \textit{builder} to the training and the test sets, according to a random or to a more realistic timestamp-based strategy. It is up to the experimenter deciding the percentage of sequences in the test set.

The \textit{recommender} module includes an abstract class that needs to be implemented by any recommender that relies on this framework. For demonstrative purposes, we have implemented four different baseline recommenders, described in Section~\ref{sec:recommenders}.

The purpose of the abstract class in the \textit{similarity} module is to compute a content-based similarity metric between two items; we have chosen to implement it as a generic cosine-based similarity.

Finally, the \textit{evaluator} module, given the test set and a recommender, computes the measures. We have decided to rely on a comprehensive set of eight metrics, which are detailed in Section~\ref{sec:metrics}.

\subsection{Recommenders}
\label{sec:recommenders}
We have included in \texttt{sequeval} four baseline recommenders, which represent an adaptation of classical non-personalized baselines to the sequence-based scenario.

\begin{description}
\item [Most Popular] The most popular recommender only considers the popularity of the items in the sequences available in the training set in order to create the recommended sequence, which is always a ranked list of the most popular items.
\item [Random] The random recommender simply creates sequences that include items randomly chosen from the ones available in the training sequences.
\item [Unigram] The unigram recommender generates sequences that contain items sampled with a probability proportional to the number of times they were observed in the training set.
\item [Bigram] The bigram recommender estimates the 1\textit{-st} order transition probabilities among all possible pair of items available in the training sequences and it recommends sequences according to those probabilities.
\end{description}

\subsection{Evaluation Metrics}
\label{sec:metrics}
In order to provide a comprehensive analysis of the recommended sequences, we have included in the \textit{evaluator} module eight metrics. We decided to consider traditional metrics like coverage and precision, and also less common ones like novelty, diversity, and serendipity. We have also introduced the metric of perplexity, because it was created for evaluating sequences~\cite{Bengio2003}.

For each sequence in the test set, a sequence of a certain length is generated by the chosen recommender, considering the same target user and the first item of the test sequence as the initial seed. The length of the sequences is a parameter of the experiment.

In the following, we briefly describe the metrics available.

\begin{description}
\item [Coverage] It expresses the items that the sequence-based RS is capable of suggesting when generating sequences similar to the ones available in the test set.
\item [Precision] It represents the fraction of recommended items that are included in the corresponding test sequence.
\item [nDPM] It indicates if the items in the recommended sequences are in the same order of the ones in the test sequences.
\item [Diversity] This metric exploits the cosine similarity in order to check if the items in the recommended sequences are enough diverse among them.
\item [Novelty] It rewards algorithms that are capable of suggesting sequences that contain items that are probably unknown to the user because they are not popular.
\item [Serendipity] It is similar to the metric of precision, but it is computed always considering popular items as irrelevant. 
\item [Confidence] It expresses how sure the recommender is of the generated sequences and it is computed as the average probability that the suggested items are correct.
\item [Perplexity] This metric can be interpreted as the number of items from which a random recommender should choose in order to obtain a similar sequence.
\end{description}

%% file: sections/03_usage.tex
\section{Demonstration and Usage}

\begin{figure}
\includegraphics[width=\linewidth]{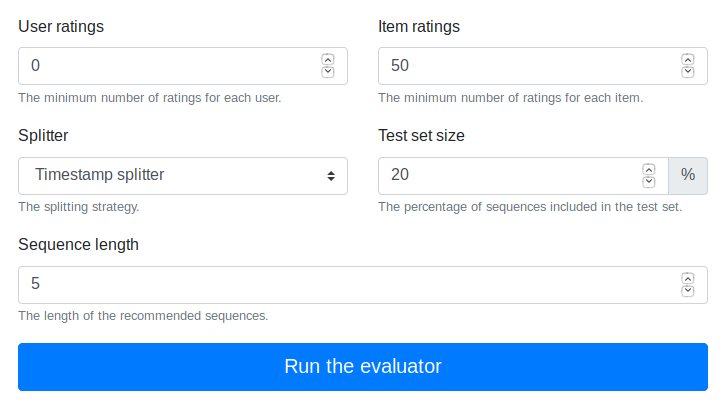}
\caption{Screenshot of the web-based interface}
\label{fig:screenshot}
\end{figure}

In order to exploit the proposed framework, it is necessary to create an implementation of the abstract recommender that must be capable, given the user and the current item of the sequence, of predicting the probabilities for all the possible items of being the next one inside the recommended sequence. It is then possible to write an evaluation script that relies on this library.

For demonstrative purposes, we have included in \texttt{sequeval} a down-sampled version of the playlists dataset originally collected by Shuo Chen from the \textit{Yes.com} website~\cite{Chen2012}.

We also provide a simple evaluation script which can be used to perform different experiments with the baseline recommenders. This script can be easily modified in order to accommodate novel recommandation techniques.

Furthermore, we have realized a web-based version of the same evaluation script in order to provide a more convenient graphical interface to perform the experiments, as shown in Figure~\ref{fig:screenshot}.

%% file: sections/04_conclusion.tex
\section{Conclusion and Future Work}
In this paper, we presented \texttt{sequeval}, a software tool for evaluating sequence-based recommender systems. The strategic choice of publishing it as an open source project fosters reuse and extension. As future work, we plan to include in \texttt{sequeval} further datasets and more complex recommendation techniques.